\documentclass[twocolumn,aps,superscriptaddress,nofootinbib]{revtex4}
\usepackage[dvips]{graphicx}
\usepackage{latexsym,amssymb,amsmath}
\usepackage{color}
\usepackage{bm}
\usepackage{enumerate}
\usepackage[bookmarksnumbered,bookmarksopen,colorlinks,citecolor=red,linkcolor=blue]{hyperref}
\usepackage{mathrsfs}
\usepackage{times}
\usepackage{csquotes}
\usepackage{multirow}
\usepackage[dvipsnames]{xcolor}

\graphicspath{{./figs/}}

\begin{document}   

\title{Resonant screening in dense and magnetized QCD matter}
\author{Guojun Huang}
\affiliation{Department of Physics, Tsinghua University, Beijing 100084, China}
\author{Jiaxing Zhao}
\affiliation{SUBATECH, Universit\'e de Nantes, IMT Atlantique, IN2P3/CNRS, 4 rue Alfred Kastler, 44307 Nantes cedex 3, France}
\author{Pengfei Zhuang}
\affiliation{Department of Physics, Tsinghua University, Beijing 100084, China}
\date{\today}

\begin{abstract}
We calculate the Debye screening mass in thermal, dense and magnetized QCD matter in the frame of resummed perturbation theory. In the limit of zero temperature, when the Landau energy level and Fermi surface of quarks match each other $\mu_q^2=2n|qB|$, where $q$, $\mu_q$ and $B$ are respectively the quark electric charge, chemical potential and external magnetic field, the screening mass diverges and the system is in the state of weakly interacting parton gas, which is very different from the known result of strongly interacting quark-gluon plasma at high temperature. The divergence disappears in thermal medium, but the screening mass oscillates with clear peaks at the matched magnetic field.    
\end{abstract}
\maketitle

\emph{Introduction.--} An immediate consequence of quarks being fermions is that quarks must satisfy the Pauli exclusion principle, which states that no two quarks can occupy the same state. The application of this pure quantum effect in quark matter at finite baryon density means a boundary separating occupied and unoccupied states in momentum space, it is called the Fermi surface controlled by the baryon chemical potential. The sharp Fermi surface leads to the expectation that there should a first-order phase transition in Quantum ChromoDynamics (QCD) systems at high baryon density. While lattice QCD simulations meet the sign problem, many effective models obtain the first-order phase transition for the chiral symmetry restoration at high density~\cite{Stephanov:2004wx,Fukushima:2013rx,Braun-Munzinger:2008szb}. A hot topic in the experimental and theoretic study of QCD thermodynamics in relativistic heavy ion collisions is the search for the critical point connecting the crossover at high temperature and the first-order transition at high baryon density~\cite{Braun-Munzinger:2008szb,Bzdak:2019pkr}.      

A strong external electromagnetic field can also induce QCD phase transitions and change the QCD medium properties, like magnetic catalysis~\cite{Bali:2012zg,Shovkovy:2012zn}, inverse magnetic catalysis~\cite{Bali:2012zg,Bruckmann:2013oba} and chiral magnetic effect~\cite{Kharzeev:2007jp,Fukushima:2008xe}. It is widely believed that the strongest electromagnetic field in nature can be created in non-central nuclear collisions~\cite{Kharzeev:2007jp,Skokov:2009qp,Voronyuk:2011jd,Deng:2012pc,Tuchin:2013ie,Wang:2021oqq,Yan:2021zjc,Chen:2021nxs}. A familiar quantum phenomenon for a fermion moving in an external magnetic field is the Landau energy level: The fermion's moving in the transverse plane perpendicular to the magnetic field is like a harmonic oscillator which leads to a discrete transverse energy~\cite{Sakurai}. A question we ask ourselves is what will happen when the sharp Fermi surface meets the discrete Landau levels. The related physics systems with high baryon density and strong magnetic field might be realized in compact stars and intermediate energy nuclear collisions~\cite{Harding:2006qn,Bzdak:2019pkr,STAR:2017sal}.      

We study in this work the color screening in thermal, dense and magnetized QCD matter. Analog to the Debye screening in electrodynamics, the color interaction between a pair of quarks is screened by the surrounding quarks and gluons~\cite{QFT}. The Debye mass $m_D$ or the Debye length $r_D\sim 1/m_D$ can be used to effectively describe the QCD phase transition: When the Debye length is shorter than the hadron averaged radius, the constituents inside the hadron cannot see each other through the color interaction and the hadron as a whole particle disappears. In finite temperature field theory, the Debye mass is defined as the static and long wavelength limit of the gluon self-energy~\cite{QFT}. In a hot QCD medium, the Hard Thermal Loop (HTL) method gives a temperature-dependent Debye screening mass $m_D(T)$~\cite{Laine:2006ns}. For a dense QCD medium, the Hard Dense Loop (HDL) provides a similar baryon chemical potential dependence $m_D(\mu_B)$~\cite{QFT,Manuel:1995td,Huang:2021ysc}. Recently, the screening effect is investigated in a thermal and magnetized QCD matter~\cite{Huang:2022fgq}, and the calculated Debye mass $m_D(T,B)$ recovers the previously obtained limits of weak and strong magnetic fields~\cite{Hasan:2020iwa,Karmakar:2018aig,Singh:2017nfa}.  

The paper is organized as follows. We first derive the Debye screening mass $m_D(T,\mu_B,B)$ at finite temperature, baryon chemical potential and magnetic field in the frame of loop-resumed perturbation theory, and compare it with the previously obtained limits of dilute quark gas, weak and strong magnetic field. We then focus on the case at zero temperature and analyze the new phenomenon induced by the quantum Fermi surface and Landau levels. We will show the numerically calculated Debye mass at zero and low temperature to see clearly the new phenomenon. We summarize the work in the end. 

\emph{Screening mass.--} The magnetic field breaks down the translation invariance, which leads to the separation of the quark momentum $p$ into a longitudinal and a transverse part $p_{||}$ and $p_\perp$, parallel and perpendicular to the magnetic field. Using the Schwinger propagator for massless quarks with electric charge $q$~\cite{Schwinger:1951nm},
\begin{eqnarray}
\label{G}
G(p) &=& -\int_0^\infty {dv\over |qB|}\Big[(\gamma\cdot p)_{||}\left(1-i{\rm sgn}(q)\gamma_1\gamma_2\tanh v\right)\nonumber\\
&&-{(\gamma\cdot p)_\perp\over \cosh^2 v}\Big]e^{{v\over |qB|}\left(p_{||}^2-{\tanh v\over v} p_\perp^2\right)},
\end{eqnarray}
where the magnetic field $B$ is explicitly shown, and the temperature $T$ and baryon chemical potential $\mu_B$ enter the calculation through the Matsubara frequency $p_0=i\omega_m=i(2m+1)\pi T$ and energy shift $p_0\to p_0-\mu_q$ with quark chemical potential $\mu_q=\mu_B/3$, one can calculate the gluon polarization induced by a quark loop, 
\begin{equation}
\label{Pi}
\Pi_{\mu\nu}(k)=g^2\int{d^4p\over (2\pi)^4}\text{Tr} \left[\gamma_\mu G(p)\gamma_\nu G(p-k)\right],
\end{equation}
where $k$ is the gluon momentum, and the quark momentum integration includes a three-momentum integration and a Matsubara frequency summation. Note that, we have neglected the Schwinger phase factor in the propagator (\ref{G}), since in the calculation of the polarization the two-phase factors for the quark and anti-quark cancel to each other. Here we have also fixed the quark flavor, we will consider the flavor summation in the end. 

Taking the usually used summation over quark loops on a chain, one can derive a non-perturbative gluon propagator~\cite{QFT}, and the Debye mass is defined as the pole of the propagator. In the cases of high temperature and/or high density, the screening mass is determined only by the polarization in the limit of zero gluon momentum $\Pi_{\mu\nu}(k_0=0,{\bm k}\to 0)$, called as HTL and HDL approximations~\cite{QFT}. Since in these cases, the polarization depends only on the external parameters, one can explicitly express the dependence as $\Pi_{\mu\nu}(T,\mu_{B},B)$. It is easy to see that all the off-diagonal elements ($\mu\neq\nu$) of the polarization vanish automatically, one needs to consider the diagonal elements only. One further divides the diagonal polarization into a parallel and a perpendicular part $\Pi_{\mu\mu}^\|$ with $\mu\in\{0,3\}$ and $\Pi_{\mu\mu}^\perp$ with $\mu\in\{1,2\}$, only the parallel part is related to the color screening mass~\cite{QFT}. Including the gluon-loop and ghost-loop contribution $\overline\Pi_{\mu\nu}$ to the gluon polarization which is only temperature dependent, since gluons and ghosts do not carry charge and chemical potential, the Debye screening mass is expressed as
\begin{equation}
\label{mD}
m_D^2(T,\mu_B,B)=-\Pi_{00}^\|(T,\mu_B,B)-\overline\Pi_{00}^\|(T).
\end{equation}    
Since the pure temperature and density dependence in the absence of magnetic field is known~\cite{Laine:2006ns,Manuel:1995td,Huang:2021ysc}, 
\begin{equation}
\label{noB}
m_D^2(T,\mu_B,0) = g^2\left[\left({N_c\over 3}+{N_f\over 6}\right)T^2+N_f{\mu_q^2\over 2\pi^2}\right]
\end{equation}
with numbers $N_c$ and $N_f$ of color and flavor degrees of freedom, we focus in the following on the shift of the squared Debye screening mass induced by $B$. From the calculation at zero baryon density~\cite{Huang:2022fgq}, it is 
\begin{eqnarray}
\label{dmD1}
\delta m_D^2(T,\mu_B,B) &=& m_D^2(T,\mu_B,B)-m_D^2(T,\mu_B,0)\nonumber\\
&=& 2g^2T\sum_{p_z,m}\bar\epsilon_-^2{\mathcal K}\left(\bar\epsilon_+^2\right),
\end{eqnarray}
where the chemical potential is included in the dimensionless variables $\bar\epsilon_\pm^2=\left[p_z^2\pm (\omega_m+i\mu_q)^2\right]/(2|qB|)$, and the function ${\mathcal K}$ is defined as ${\mathcal K}(x)=1/(2x^2)+1/x-\psi'(x)$ with $\psi(x)=\Gamma'(x)/\Gamma(x)$ controlled by the Gamma function. The quark frequency summation $\sum_{m=-\infty}^\infty$ and the longitudinal momentum integration $\sum_{p_z}=\int dp_z/(2\pi)^2$ are explicitly shown in (\ref{dmD1}), and the summation over Landau levels for the transverse momentum $p_n=\sqrt{2n|qB|}$ is reflected in the recurrence relation of the function ${\mathcal K}$,    
\begin{eqnarray}
\label{K1}
{\mathcal K}(x) &=& 1/(2x^2)+1/x-1/(2(x+N)^2)-1/(x+N)\nonumber\\
&&+{\mathcal K}(x+N)-\sum_{n=0}^{N-1}1/(x+n)^2
\end{eqnarray}
with  
\begin{equation}
\label{N1}
N =\left[\textrm{Floor}\left({\mu_q^2-\pi^2T^2\over 2|qB|}\right)+1\right]\theta(\mu_q-\pi T).
\end{equation}

To understand the physics of the summation here, we consider the Fermi surface for quarks moving in an external magnetic field. It is determined by the quark Fermi-energy $\epsilon_n=\sqrt{p_z^2+p_n^2} = \mu_q$ which restricts the quark momentum $p_z^2+2n|qB|\le \mu_q^2$. Since $p_z^2$ is positive, the maximum Landau level is $\text{Floor}[\mu_q^2/(2|qB|)]$. Extending the analysis to finite temperature, the restriction condition becomes $p_z^2+2n|qB|+\omega_m^2\le \mu_q^2$. Considering again the minimum longitudinal momentum $p_z=0$ and the minimum thermal energy $\pi T$, the maximum Landau level becomes $\text{Floor}(\mu_q^2-\pi^2T^2)/(2|qB|)$ under the condition of $\mu_q > \pi T$, which is expressed as $N-1$ in (\ref{K1}) and (\ref{N1}). When the Fermi surface is not high enough to overcome the thermal energy $\mu_q < \pi T$, there is $N=0$ and the summation in (\ref{K1}) vanishes automatically.  

The recurrence relation (\ref{K1}) leads to 
\begin{equation}
\label{K2}
{\mathcal K}\left(\bar\epsilon_+^2\right) ={1\over \bar\epsilon_+^2} -{1\over \bar\epsilon_N^2}-{1\over 2}\sum_{n=0}^N{\Delta^n_N\over \left(\bar\epsilon_n^2\right)^2}+{\mathcal K}\left(\bar\epsilon_N^2\right)
\end{equation}
with $\Delta^n_N=2-\delta_{0n}-\delta_{nN}$ and $\bar\epsilon_n^2=\bar\epsilon_{+}^2+n$. Substituting ${\mathcal K}$ into the mass shift (\ref{dmD1}), after a straightforward but a little tedious calculation, see the details shown in Suppl.~\ref{s1}, the total mass shift is separated into two components,
\begin{eqnarray}
\label{full}
\delta m_D^2 &=& \delta m_1^2+\delta m_2^2,\nonumber\\
\delta m_1^2 &=&-{g^2\over 4T} \sum_{p_z,s}\Big[p_z^2{\mathcal S}_N^s-{|qB|\over 2}\sum_{n=0}^N\Delta^n_N\text{sech}^2(\bar\epsilon_n^s)\Big],\nonumber\\
\delta m_2^2 &=&{2g^2T^2\over\pi^{1/2}}\int_0^\infty {d\xi\over \xi^2}e^{c\xi^2}\vartheta_2\big(a\xi^2,e^{-\xi^2}\big){\mathcal M}\left(b\xi^2\right)
\end{eqnarray}
with the constants $a^2=\mu_q^2/(4\pi^2T^2)$, $b=|qB|/(4\pi^2T^2)$, $c=a^2-2Nb$ and $\bar\epsilon_n^s=(p_{z}^{2}+p_{n}^{2}+s\mu_q)/(2T)$ and the functions 
${\mathcal S}_N^s = \textrm{sech}^2(\bar\epsilon_0^s) -\textrm{sech}^2(\bar\epsilon_N^s)$, $\vartheta_2(u,x) = 2x^{1/4}\sum_{i=0}^\infty x^{i(i+1)}\cos\left((2i+1)u\right)$ and ${\mathcal M}(x) = 1-x^2/\sinh^2 x+2Nx\left(1-x\coth x\right)$. The summation $\sum_{s=\pm}$ is over quarks and anti-quarks. Note that the first component $\delta m_1^2$ disappears automatically for $N=0$ due to ${\mathcal S}_0^s=0$ and $\Delta^n_0=0$, the often used approximation~\cite{Hattori:2015aki} of taking only the lowest Landau level is included in the second component $\delta m_2^2$.     

\emph {Weak and strong magnetic field.--} We now start to consider some physics limits of the general mass shift (\ref{full}). We first consider a dilute quark gas with $\mu_q\to 0$. In this case with $N=0$, the component $\delta m_1^2$ vanishes automatically, $\delta m_D^2(T,0,B)=\delta m_2^2(T,0,B)$ goes back to the known result in hot medium~\cite{Huang:2022fgq}. 

We then consider the limit of weak magnetic field with $|qB|\to 0$. In this limit, the second component $\delta m_2^2$ disappears due to ${\mathcal M}(0)=0$. From the derivation provided in Suppl.II the first component $\delta m_1^2$ vanishes too. Therefore, we obtain the Taylor expansion of the mass shift in terms of $|qB|$ for a weak magnetic field at any temperature and baryon chemical potential,
\begin{equation}
\label{weak}
\delta m_D^2 = {14\zeta(3)\over 9(2\pi)^4}{g^2\over T^2}|eB|^2+\mathcal O(|eB|^4),
\end{equation}
where we have taken into account the quark flavor summation. While the quark chemical potential is flavor independent, the electric charge $q$ depends on the flavor $q_u=2e/3, q_d=-e/3$ and $q_s=-e/3$ . 
 
The other limit we consider is the strong magnetic field limit with $\sqrt{|qB|}\gg T,\mu_B$. Taking a variable transformation $\xi=T\eta/|qB|^{1/2}$ and employing the limit 
\begin{equation}
\label{limit1}
\lim_{|qB|\to\infty} e^{a^2\xi^2}\vartheta_2\big(a\xi^2,e^{-\xi^2}\big)=\sqrt{\pi|qB|}/(T\eta),
\end{equation}
the component $\delta m_2^2$ becomes
\begin{equation}
\label{strong1}
\delta m_2^2 ={g^2\over 4\pi^2}|qB|\theta(\pi T-\mu_q).
\end{equation}
This covers the familiar result~\cite{Singh:2017nfa} including only the contribution from the lowest Landau level. More details about the derivation of (\ref{strong1}) are provided in Suppl.III. For the component $\delta m_1^2$, in the limit of strong magnetic field with $N=1$ at $\mu_q > \pi T$, by using the delta function
\begin{equation}
\label{delta}
\delta(x) = \lim_{T/|qB|^{1/2}\to 0}\left[{1\over 4T/|qB|^{1/2}}\textrm{sech}^2\left({x\over 2T/|qB|^{1/2}}\right)\right],
\end{equation}
we have 
\begin{equation}
\label{strong2}
\delta m_1^2 = {g^2\over 4\pi^2}\left(|qB|-2\mu_q^2-{2\over3}\pi^2T^2\right)\theta(\mu_q-\pi T).
\end{equation}

Considering the summation over all quark flavors, the total mass shift in a strong magnetic field is  
\begin{equation}
\label{strong}
\delta m_D^2 ={g^2\over 3\pi^2}|eB|-{N_fg^2\over 2\pi^2}\left(\mu_q^2+{\pi^2\over 3}T^2\right)\theta(\mu_q-\pi T).
\end{equation}
\begin{figure}[!htbp]
	\includegraphics[width=7.5cm]{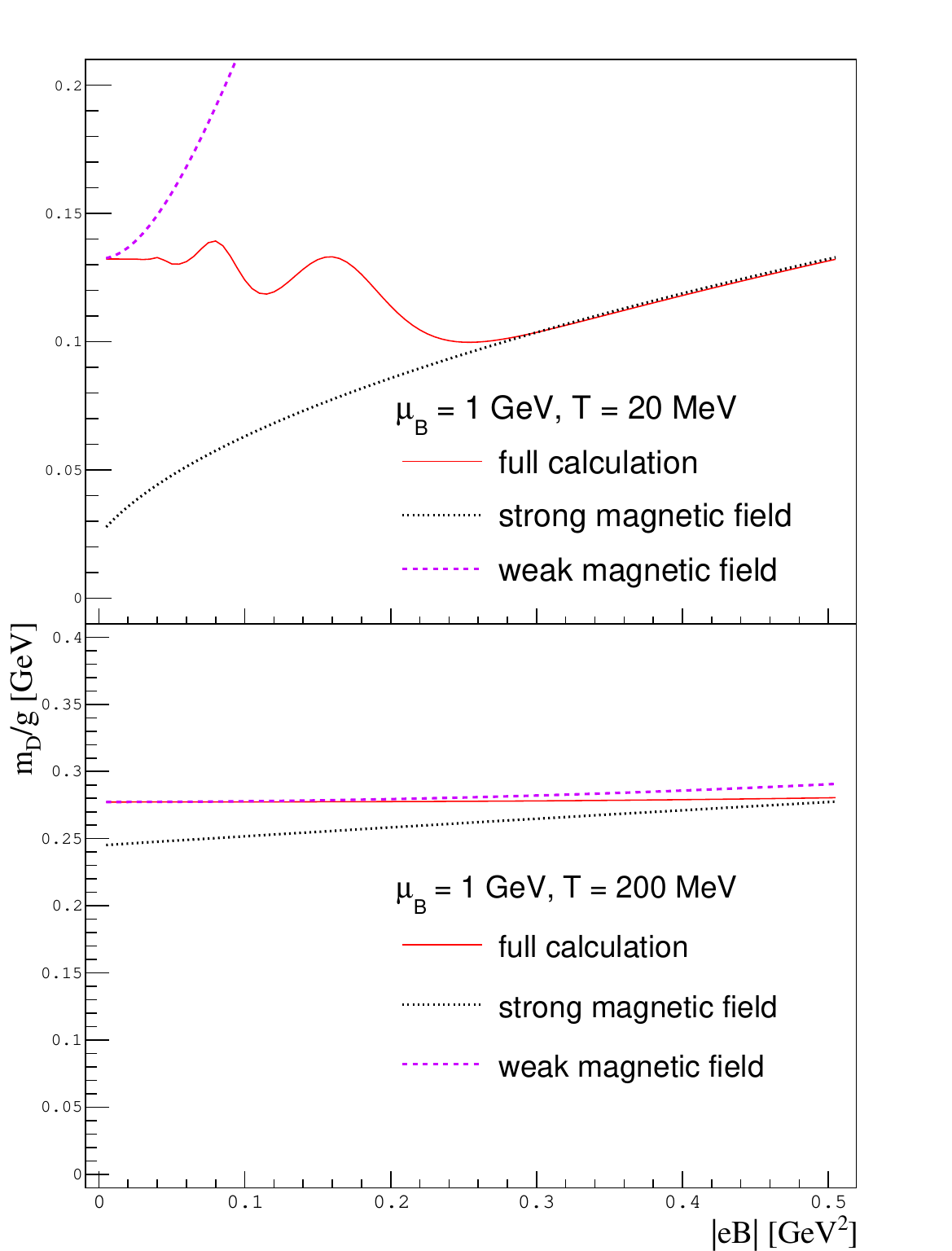}
	\caption{The scaled screening mass $m_D/g$ as a function of the strength $|eB|$ of the external magnetic field at fixed baryon chemical potential $\mu_B=1$ GeV and two temperatures $T=20$ MeV (upper panel) and $200$ MeV (lower panel). The solid, dashed and dotted lines are the full calculation and two approximations with weak and strong magnetic fields. }
	\label{fig1}
\end{figure}

The magnetic field dependence of the scaled screening mass $m_D/g$ (\ref{full}) and the comparison with the approximations of weak (\ref{weak}) and strong (\ref{strong}) magnetic field are shown in Fig.\ref{fig1} at high baryon chemical potential $\mu_B=1 $ GeV and low and high temperatures $T=20$ and $100$ MeV. In the frame of loop resummation, $m_D/g$ is no longer coupling constant dependent. The integrated function in $\delta m_2^2$ (\ref{full}) looks like divergent at $\xi=0$, but it is convergent due to the limits $\lim_{\xi\to 0}e^{a^2\xi^2}\vartheta_2(a\xi^2,e^{-\xi^2})\sim 1/\xi$ and $\lim_{\xi\to 0} M_N(b\xi^2)\sim \xi^4+\cdots$. The color and flavor numbers are chosen as $N_c=N_f=3$. Naturally, the full calculation is always in between the two approximations.  Because the external magnetic field breaks down the isotropy symmetry but random thermal motion restores the symmetry, the approximation of weak magnetic field becomes better and the approximation of strong magnetic field becomes worse with increasing temperature, as a consequence of the competition of the two effects.   

\emph{Resonant screening.--} An interesting phenomenon shown in Fig.\ref{fig1} is the oscillation behavior of the screening mass at low temperature. To understand the physics behind it we turn to discuss the zero-temperature limit of Eq.(\ref{full}). Similar to the treatment for the integration over $\xi$ in a strong magnetic field, after taking the variable transformation from $\xi$ to $\eta$, the integrated function can be expressed as a complete differential, and it is zero at both the lower and upper limit of the integration, which leads to $\delta m_2^2=0$ at $T\to 0$. By taking the limits
\begin{eqnarray}
\label{limit2}
\lim_{T\to 0}\tanh (x/T) &=& 2\theta(x)-1,\\
\partial x\left(\tanh(x/(2T)\right) &=& \textrm{sech}^2(x/(2T))/(2T)\to 2\delta(x),\nonumber
\end{eqnarray}
the total mass shift becomes
\begin{eqnarray}
\label{T0}
\delta  m_D^2 &=& \delta m_1^2\nonumber\\
&=& -{g^2\mu_q^2\over 2\pi^2}+{g^2\mu_q|qB|\over (2\pi)^2}\sum_{n=0}^{N-1}{2-\delta_{0n}\over\sqrt{\mu_q^2-2n|qB|}}
\end{eqnarray}
with the maximum Landau level $N-1=\text{Floor}[\mu_q^2/(2|qB|)]$. The details to derive (\ref{T0}) are shown in Suppl.VI. Considering the result (\ref{noB}) in the absence of magnetic field, and taking into account all the flavors, the total mass square in zero temperature limit is
\begin{equation}
m_D^2 = {g^2\mu_q\over (2\pi)^2}\sum_f\sum_{n=0}^{N-1}{|q_fB|(2-\delta_{0n})\over\sqrt{\mu_q^2-2n|q_fB|}}.
\end{equation}  

\begin{figure}[!htb]
	\includegraphics[width=8.0cm]{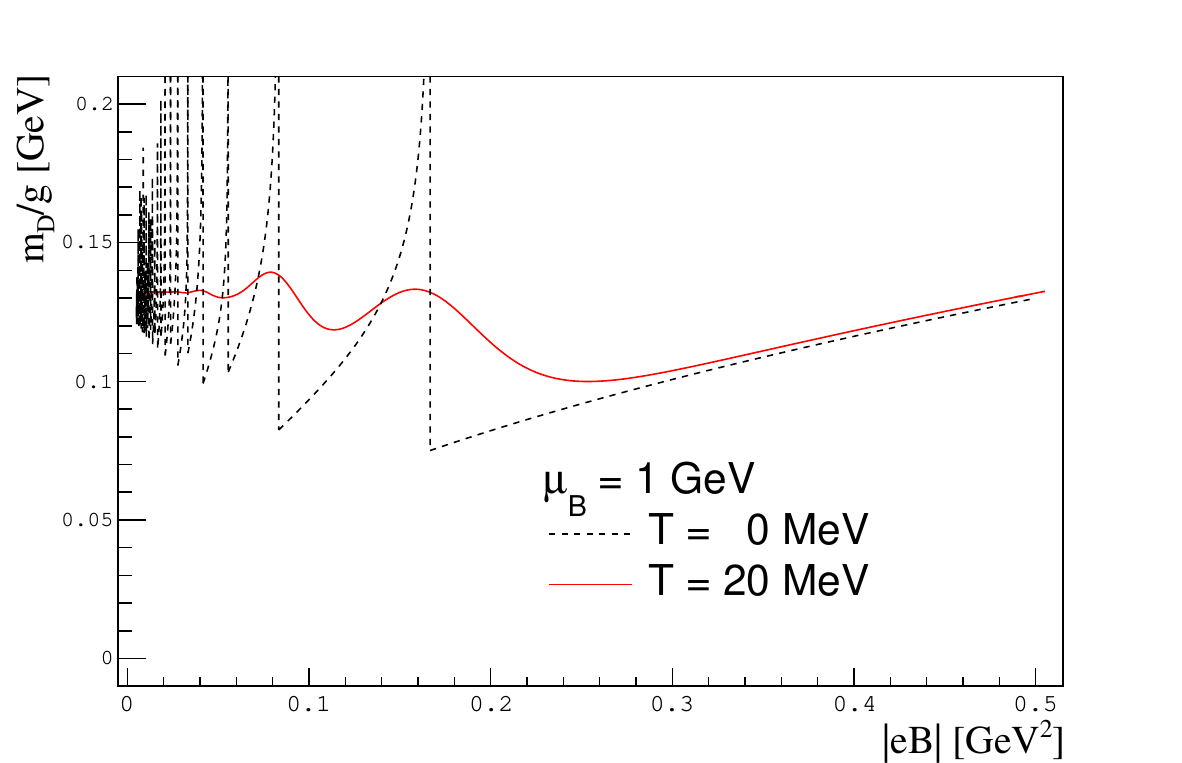}
	\caption{The scaled screening mass $m_D/g$ as a function of the strength $|eB|$ of the external magnetic field at fixed baryon chemical potential $\mu_B=1$ GeV in zero-temperature limit and at temperature $T=20$ MeV. }
	\label{fig2}
\end{figure}

When the Fermi surface matches the Landau energy level, $\mu_q^2=2n|q_fB|$, controlled by the chemical potential and magnetic field, the Debye mass diverges $m_D\to \infty$ and the screening length approaches to zero $r_D\to 0$. That means a full screening: the color interaction between a pair of quarks is completely screened, and the medium becomes a weakly interacting quark gas. This phenomenon at high baryon density is different from the known result that the QCD medium is a strongly interacting quark-gluon plasma at high temperature~\cite{Braun-Munzinger:2008szb}. The numerical calculation in the limit of zero temperature and the modification by the thermal motion are shown in Fig.\ref{fig2}. For fixed chemical potential $\mu_B=1$ GeV, the magnetic fields where the mass diverges are
$|eB|=1/6, 1/12,\cdots$ GeV$^2$ for $d$ and $s$ quarks and $|eB|=1/12, 1/24,\cdots$ GeV$^2$ for $u$ quarks, corresponding to the first, second, third and other divergences counted from the right. When the magnetic field is too strong to satisfy the matching condition, $|q_fB|>\mu_q^2/2=\mu_B^2/18$, the mass becomes chemical potential independent and is linear in the magnetic field,
\begin{equation}
m_D^2 = {g^2\over 3\pi^2}|eB|,
\end{equation} 
see the black dashed line in the region of strong magnetic field shown in Fig.\ref{fig2}.

The above divergence induced by the matched Fermi surface and Landau energy level is similar to the well-known resonant transmission in quantum mechanics~\cite{Sakurai}. For the particle tunneling through a potential well, when the particle energy and the well structure (the depth and width of the well) meet certain condition, the particle completely penetrates the well with transmission coefficient $T=1$ without any reflection $R=0$. From the comparison with this resonant transmission, we call the above divergence of the screening mass as resonant screening.      

When the thermal motion is turned on, the sharp Fermi surface becomes a smooth distribution, and the divergence is washed away. When the temperature is not so high, the thermal motion is not yet strong enough, the mass oscillates with the magnetic field, and the divergence is changed to a peak, see the red lines in Figs. (\ref{fig1}) and (\ref{fig2}) at $T=20$ MeV. 

{\emph{Summary.--} We calculated in this paper the Debye screening mass in QCD matter at finite temperature, baryon density and magnetic field in the resummed perturbation theory. The Landau energy levels of moving quarks in the external magnetic field are explicitly shown in the mass. Three limits of the general result, namely the dilute quark gas, and weak and strong magnetic field, are discussed in detail. We focused on the dense and magnetized medium in the limit of vanishing temperature. In this case, when the Landau energy level matches the Fermi surface of quarks, the screening mass goes to infinity, indicating that the color interaction between a pair of quarks is completely screened. We call this full screening as resonant screening in comparison with the resonant transmission in quantum mechanics, the location of the resonance is determined by the matching condition $\mu_q^2=2n|qB|$. While the random thermal motion smears the divergence, there is still oscillation with clear peaks at low temperatures. The resonant screening or screening peaks may have applications in dense and magnetized matter created in compact stars and intermediate energy nuclear collisions.  

\vspace{1cm}
\noindent {\bf Acknowledgement}: The work is supported by the NSFC grants Nos. 12247185, 11890712, and 12075129, the Guangdong Major Project of Basic and Applied Basic Research No. 2020B0301030008, the funding from the European Union’s Horizon 2020 research, and innovation program under grant agreement No. 824093 (STRONG-2020).


\begin{thebibliography}{20}

\bibitem{Stephanov:2004wx}
M.~A.~Stephanov,
Prog. Theor. Phys. Suppl. \textbf{153}, 139-156 (2004).

\bibitem{Fukushima:2013rx}
K.~Fukushima and C.~Sasaki,
Prog. Part. Nucl. Phys. \textbf{72}, 99-154 (2013).

\bibitem{Braun-Munzinger:2008szb}
P.~Braun-Munzinger and J.~Wambach,
Rev. Mod. Phys. \textbf{81}, 1031-1050 (2009).


\bibitem{Bzdak:2019pkr}
A.~Bzdak, S.~Esumi, V.~Koch, J.~Liao, M.~Stephanov and N.~Xu,
Phys. Rept. \textbf{853}, 1-87 (2020).


\bibitem{Bali:2012zg}
G.~S.~Bali, F.~Bruckmann, G.~Endrodi, Z.~Fodor, S.~D.~Katz and A.~Schafer,
Phys. Rev. D \textbf{86}, 071502 (2012).

\bibitem{Shovkovy:2012zn}
I.~A.~Shovkovy,
Lect. Notes Phys. \textbf{871}, 13-49 (2013).

\bibitem{Bruckmann:2013oba}
F.~Bruckmann, G.~Endrodi and T.~G.~Kovacs,
JHEP \textbf{04}, 112 (2013).

\bibitem{Kharzeev:2007jp}
D.~E.~Kharzeev, L.~D.~McLerran and H.~J.~Warringa,
Nucl. Phys. A \textbf{803}, 227-253 (2008).


\bibitem{Fukushima:2008xe}
K.~Fukushima, D.~E.~Kharzeev and H.~J.~Warringa,
Phys. Rev. D \textbf{78}, 074033 (2008).


\bibitem{Skokov:2009qp}
V.~Skokov, A.~Y.~Illarionov and V.~Toneev,
Int. J. Mod. Phys. A \textbf{24}, 5925-5932 (2009).


\bibitem{Voronyuk:2011jd}
V.~Voronyuk, V.~D.~Toneev, W.~Cassing, E.~L.~Bratkovskaya, V.~P.~Konchakovski and S.~A.~Voloshin,
Phys. Rev. C \textbf{83}, 054911 (2011).

\bibitem{Deng:2012pc}
W.~T.~Deng and X.~G.~Huang,
Phys. Rev. C \textbf{85}, 044907 (2012).

\bibitem{Tuchin:2013ie}
K.~Tuchin,
Adv. High Energy Phys. \textbf{2013}, 490495 (2013).

\bibitem{Wang:2021oqq}
Z.~Wang, J.~Zhao, C.~Greiner, Z.~Xu and P.~Zhuang,
Phys. Rev. C \textbf{105}, no.4, L041901 (2022).

\bibitem{Yan:2021zjc}
L.~Yan and X.~G.~Huang,
Phys. Rev. D \textbf{107}, no.9, 094028 (2023).

\bibitem{Chen:2021nxs}
Y.~Chen, X.~L.~Sheng and G.~L.~Ma,
Nucl. Phys. A \textbf{1011}, 122199 (2021).

\bibitem{Sakurai}
J.~J.~Sakurai, Modern Quantum Mechanics, Revised Edition, 1994, Addison-Wesley Publishing Company.

\bibitem{Harding:2006qn}
A.~K.~Harding and D.~Lai,
Rept. Prog. Phys. \textbf{69}, 2631 (2006).

\bibitem{STAR:2017sal}
L.~Adamczyk \textit{et al.} [STAR],
Phys. Rev. C \textbf{96}, no.4, 044904 (2017).

\bibitem{QFT}
J.~I.~Kapusta and C.~Gale, Finite-temperature field theory: Principles and applications (Cambridge University Press, Cambridge, 2009).
M.~Le~Bellac, Thermal Field Theory (Cambridge University Press, Cambridge, 1996).

\bibitem{Laine:2006ns}
M.~Laine, O.~Philipsen, P.~Romatschke and M.~Tassler,
JHEP \textbf{03}, 054 (2007).


\bibitem{Manuel:1995td}
C.~Manuel,
Phys. Rev. D \textbf{53}, 5866-5873 (1996).

\bibitem{Huang:2021ysc}
G.~Huang and P.~Zhuang,
Phys. Rev. D \textbf{104}, no.7, 074001 (2021).

\bibitem{Huang:2022fgq}
G.~Huang, J.~Zhao and P.~Zhuang,
[arXiv:2208.01407 [hep-ph]].

\bibitem{Hasan:2020iwa}
M.~Hasan and B.~K.~Patra,
Phys. Rev. D \textbf{102}, no.3, 036020 (2020).

\bibitem{Karmakar:2018aig}
B.~Karmakar, A.~Bandyopadhyay, N.~Haque and M.~G.~Mustafa,
Eur. Phys. J. C \textbf{79}, no.8, 658 (2019).

\bibitem{Singh:2017nfa}
B.~Singh, L.~Thakur and H.~Mishra,
Phys. Rev. D \textbf{97}, no.9, 096011 (2018).


\bibitem{Schwinger:1951nm}
J.~S.~Schwinger,
Phys. Rev. \textbf{82}, 664-679 (1951).

\bibitem{Hattori:2015aki}
K.~Hattori, T.~Kojo and N.~Su,
Nucl. Phys. A \textbf{951}, 1-30 (2016).

\end{thebibliography}

\clearpage
\begin{widetext}
\begin{center}
\textbf{\large Supplemental Materials}
\end{center}

We provide supplemental materials in the sequence according to the contents in the main material. We provide in Sections~\ref{s1}, \ref{s2} and \ref{s3} the details to derive the full mass shift $\delta m_D^2(T,\mu_B,B)$ (\ref{full}) in general case with nonzero temperature, baryon density and magnetic field and its approximations (\ref{weak}) and (\ref{strong}) in weak and strong magnetic field. We focus on the screening mass (\ref{T0}) in dense and magnetized matter at zero temperature in Section \ref{s4}. 

\section{Screening Mass in General Case}
\label{s1}
Substituting the recurrence relation (\ref{K2}) into the mass shift (\ref{dmD1}), we have 
\begin{equation}
\label{s11}
\delta m_D^2 = 2g^2T\sum_{p_z,m}\bar\epsilon_-^2\left[{1\over \bar\epsilon_+^2} -{1\over \bar\epsilon_N^2}-{1\over 2}\sum_{n=0}^N{\Delta^n_N\over \left(\bar\epsilon_n^2\right)^2}+{\mathcal K}\left(\bar\epsilon_N^2\right)\right].
\end{equation}
After the Matsubara frequency summation and longitudinal momentum integration by parts, the first two terms becomes
\begin{eqnarray}
\label{s12}
2T\sum_{p_z,m} \bar\epsilon_-^2\left({1\over \bar\epsilon_+^2}-{1\over \bar\epsilon_N^2}\right) &=& \sum_{p_z,s,s'}{\partial\over \partial p_z}\left[p_z \left(\epsilon_N+s'p_z\right)/4\right]\left[s'\tanh\left(\overline\epsilon_0^s\right)-\tanh\left(\overline\epsilon_N^s\right)\right]\nonumber\\
&=&-{2N|qB|\over (2\pi)^2}-{1\over 4T}\sum_{p_z,s}p_z^2{\mathcal S}_N^s.
\end{eqnarray} 
For the third term, doing again the frequency summation and momentum integration by parts, it becomes
\begin{eqnarray}
\label{s13}
-T\sum_{p_z,m}\sum_{n=0}^N\Delta_N^n{\bar\epsilon_-^2\over\left(\bar\epsilon_n^2\right)^2} &=& |qB|\sum_{n=0}^N\Delta_N^n\left(1+{1\over 8T}\sum_{p_z,s}\text{sech}^2\left(\bar\epsilon_n^s\right)\right)\nonumber\\ 
&=& {2N|qB|\over (2\pi)^2}+{|qB|\over 8T}\sum_{p_z,s}\sum_{n=0}^N\Delta_N^n\text{sech}^2\left(\bar\epsilon_n^s\right),
\end{eqnarray} 
where we have used the summation
\begin{equation}
\sum_{n=0}^N\Delta_N^n=2N.
\end{equation}
By using the integral representation for ${\mathcal K}$,
\begin{equation}
{\mathcal K}(x) = 4b\int_0^\infty d\xi\ e^{-2bx\xi^2}\xi\left[1-b\xi^2\coth\left(b\xi^2\right)\right],
\end{equation}
doing the $\xi$-integration by parts, and then taking the frequency summation 
\begin{equation}
\sum_m e^{-2b\bar\epsilon_N^2\xi^{2}}=e^{\left(c-p_z^2/(4\pi^2T^2)\right)\xi^2}\vartheta_2\left(a\xi^2,e^{-\xi^2}\right)
\end{equation}
and momentum integration, the last term in (\ref{s11}) is exactly the second component $\delta m_2^2$. Considering the contributions from all the four terms, we obtain the total mass shift shown in (\ref{full}).

\section{Weak Magnetic Field}
\label{s2}
In this Section we prove that the first component $\delta m_1^2$ disappears in the limit of weak magnetic field, namely our calculation in general case goes back to the usually known result without magnetic field. By taking into account the limit
\begin{equation}
\label{N2}
\lim_{|qB|\to 0} 2N|qB| = \left(\mu_q^2-\pi^2T^2\right)\theta(\mu_q-\pi T)
\end{equation}
and the change from Landau level summation $\sum_n$ into Riemann integration by using the transformation $p_n^2=2n|qB|\to\zeta$ in the definition of ${\mathcal S}_{\bar\mu^2}^s$ and $\bar\epsilon_\zeta^s$, the first component becomes 
\begin{equation}
\label{dmD3}
\delta m_1^2 = \theta(\mu_q-\pi T){g^2\over 8T} \sum_{p_z,s} \left[-2p_z^2{\mathcal S}_{\bar\mu^2}^s+\int_0^{\bar\mu^2}d\zeta\ \text{sech}^2(\bar\epsilon_\zeta^s)\right]
\end{equation}
with $\bar\mu^2=\mu_q^2-\pi^2 T^2$. By writing the function $\text{sech}^2(\bar\epsilon_\zeta^s)$ in terms of two partial differentials,   
\begin{equation}
\label{diff}
\text{sech}^2\left(\bar\epsilon_\zeta^s\right) = \partial_{p_z}\left[p_z\ \text{sech}^2\left(\bar\epsilon_\zeta^s\right)\right]-2p_z^2\partial_\zeta\text{sech}^2\left(\bar\epsilon_\zeta^s\right),
\end{equation}
the integration of the first term over $p_z$ vanishes, and the integration of the second term over $\zeta$ cancels exactly the first term with ${\mathcal S}$ in Eq.(\ref{dmD3}). Therefore, the component $\delta m_1^2$ and in turn the total mass shift $\delta m_D^2$ disappear in the limit of weak magnetic field. 

\section{Strong Magnetic Field}
\label{s3}
In strong magnetic field, the maximum Landau level is a small integral and the lowest Landau level becomes the dominant contribution to the QCD thermodynamics. Taking the transformation $\xi=T\eta/{|qB|^{1/2}}$ and employing the limit (\ref{limit1}), the component $\delta m_2^2$ becomes
\begin{eqnarray}
\label{dmD22}
\delta m_2^2 &=& 2g^2|qB|\int_0^\infty d\eta\ {1\over \eta^3}e^{-{N\over 2\pi^2}\eta^2}{\mathcal M}_N\left({\eta^2\over 4\pi^2}\right)\nonumber\\
&=& {g^2|qB|\over 4\pi^2}\int_0^\infty d\eta {\partial\over\partial\eta}\left[e^{-{N\over 2\pi^2}\eta^2}\left(\coth\left({\eta^{2}\over4\pi^{2}}\right)-\left({\eta^{2}\over4\pi^{2}}\right)^{-1}\right)\right]\nonumber\\
&=& {g^2|qB|\over 4\pi^2}\delta_{0N}
\end{eqnarray}
which is zero unless $N=0$ at $\mu_q <\pi T$.

\section{Resonant Screening}
\label{s4}
In the limit of zero temperature, the QCD thermodynamics is controlled by the sharp Fermi surface, and the calculation is largely simplified. By taking the limits (\ref{limit2}), the total mass shift becomes
\begin{equation}
\delta  m_D^2 =-g^2\sum_{p_z,s}\Big[p_z^2\left(\delta\left(\epsilon_0+s\mu_q\right)-\delta\left(\epsilon_N+s\mu_q\right)\right) -{|qB|\over 2}\sum_{n=0}^N\Delta_N^n\delta\left(\epsilon_n+s\mu_q\right)\Big].
\end{equation}
Taking the integration over the momentum and summation over quarks and anti-quarks, it becomes the result (\ref{T0}) showing the resonant screening. 

\end{widetext}

\end{document}